\documentclass{aa}
\usepackage{graphicx}
\usepackage{natbib}
\bibpunct{(}{)}{;}{a}{}{,}

\begin{document}

\title{Spectral evolution of star clusters in the Large Magellanic
  Cloud}

\subtitle{I. Blue concentrated clusters in the age range 40-300\,Myr}

\author{J. F. C. Santos Jr.\inst{1}, J. J. 
Clari\'a\inst{2}\thanks{Visiting Astronomer, Complejo Astron\'omico 
El Leoncito operated under 
agreement between the Consejo Nacional de Investigaciones Cient\'{\i}ficas y 
T\'ecnicas de la Rep\'ublica Argentina and the National Universities of La 
Plata, C\'ordoba and San Juan}, A. V. Ahumada\inst{2}$^{\star}$,\\
E. Bica\inst{3}, A. E. Piatti\inst{4}, M. C. Parisi\inst{2}$^{\star}$ }

\institute{Departamento de F\'{\i}sica, ICEx, UFMG,
CP 702, 30123-970, Belo Horizonte, MG, Brazil \and Observatorio 
Astron\'omico, Laprida 854, 5000, C\'ordoba, Argentina \and 
Depto. de Astronomia, UFRGS, CP 15051, 91500-970, Porto Alegre, Brazil
\and Instituto de Astronom\'{\i}a y F\'{\i}sica del Espacio, CC 67, Suc.
28, 1428, Buenos Aires, Argentina}

\offprints{J. F. C. Santos Jr., \email{jsantos@fisica.ufmg.br}}

\date{Received / Accepted }

 
  \abstract
{}
{Integrated spectroscopy of a sample of 
17 blue concentrated Large Magellanic Cloud (LMC) clusters 
is presented and its spectral evolution studied. 
The spectra span the range $\approx$ (3600-6800)\,\AA~ 
with a resolution of $\approx$ 14\,\AA~ FWHM,
being used to determine cluster ages and, in connection with
their spatial distribution, to explore the LMC structure
and cluster formation history. 
} 
{Cluster reddening 
values were estimated by interpolation, using the available 
extinction maps.
We used two methods to derive cluster ages: (i)~{\it template matching}, 
in which 
line strengths and continuum distribution of the cluster spectra were compared
and matched to those of template clusters with known astrophysical properties, and 
(ii)~{\it equivalent width} (EW) method, in which new age/metallicity 
calibrations were used together
with diagnostic diagrams involving the sum of EWs of selected  spectral lines 
(K\,Ca\,II, G band (CH), Mg\,I, H$\delta$, 
H$\gamma$ and  H$\beta$).}
{
The derived 
cluster ages range from 40\,Myr ({NGC\,2130}
and {SL\,237}) to 300\,Myr ({NGC\,1932} and {SL\,709}),
a good agreement between the results of the two methods being obtained.
Combining the present sample with additional ones indicates that
cluster deprojected distances from the LMC center are related to 
age in the sense that inner clusters tend to be younger.
}
{Spectral libraries of star clusters are useful datasets 
for spectral classifications and extraction of parameter information for target
star clusters and galaxies. The present cluster sample complements 
previous ones, in an effort to gather
a spectral library with several clusters per age bin.}


\authorrunning{J.F.C. Santos Jr. et al.}
\titlerunning{Spectral evolution of LMC clusters}

\maketitle

\section{Introduction}

Star clusters in the Magellanic Clouds can help us to have a better insight 
about the star formation history of the galaxies as a whole 
\citep[][]{m98,psc01,psg02}. In particular, the large number
of young and intermediate-age star clusters in the Large Magellanic Cloud 
(LMC) composes an ensemble which allows to probe the galaxy structure and 
evolution during the
last few Gyrs. Indeed, the total estimated number of clusters in the LMC 
is $\approx4200$ \citep{h88}, which compares to the catalogued 4089
clusters, associations with characteristics of somewhat looser clusters 
and newly-formed ones \citep{bsdo99}.
Episodes of enhancement or reduction in cluster formation
can be traced back in the galaxy's history if statistically meaningful 
samples of clusters with well-determined properties are considered. 

The analysis of continuum and absorption lines of the integrated spectra
of star clusters has provided accurate estimates of their ages and 
metallicities.
The determination of these properties for distant and/or compact clusters,
in which individual stars are not observable, needs a calibration based
on clusters of well-known ages and metallicities. These
age and metallicity calibrations are established by means of equivalent width
(EW) measurements, in particular for conspicuous lines of H, Fe and Mg
in the visible spectra. Based on these data it is possible to 
characterize clusters
of unknown properties as well as to use the integrated spectra as 
reference populations (templates) to investigate more complex stellar 
systems. In order to accomplish these two aims, integrated spectra of 
star clusters in the 
Magellanic Clouds \citep[e.g. ][]{bas90,setal95,psc05} and in the Galaxy 
\citep[e.g. ][]{sb93,pbc02} have been observed, catalogued 
\citep{sab02}, and 
analyzed \citep[][ hereafter SP]{sp04}.

The objective of the present work is to derive age for a sample of 
17 LMC clusters, all of type II in the sequence defined by
\citet[][ hereafter SWB]{swb80}, which corresponds 
to the age range (30-70\,Myr) 
according to posterior photometric calibrations \citep[e.g. ][]{bcdsp}.
The compact nature and high surface brightness of the present sample 
clusters make them good targets for integrated spectroscopy.
For such young clusters we derived ages but it was not possible to 
derive metallicities using the proposed methods.  Even though, cluster 
ages and deprojected positions provide useful information on the LMC 
properties. When our age estimates are combined with a 
larger sample \citep{pbg03} with ages in the same scale 
as in the present work, the interpretation of the data regarding
the LMC structure and cluster formation history is improved.

The present cluster sample complements previous ones, in an attempt to provide 
a spectral library with several clusters per age bin. At the same time, we 
study the clusters themselves individually, determining their ages and 
analyzing their spatial distribution, in order to explore the LMC structure
and cluster formation history.
To estimate the clusters' ages, we employ the 
new calibrations and diagnostic diagrams recently provided by 
SP for visible integrated spectra, along with 
template spectra \citep[e.g.][]{setal95,aetal02}. The first method 
is based on the EW of Balmer H, Mg and Fe lines and the
second one on the continuum distribution of clusters with well-known 
properties. We confirm the reliability 
of the procedure proposed by SP to determine clusters' ages, since we 
included in the sample not only unstudied or poorly studied clusters, but also 
the cluster {NGC\,1839} which has been previously observed by means of
Washington photometry \citep{pgbc03}.

\section{Observations}

The observations were carried out at the Complejo Astron\'omico El Leoncito
(CASLEO, San Juan, Argentina) with the
2.15\,m telescope during four nights in December 2003. We employed a 
CCD camera 
containing a Tektroniks chip of 1024 x 1024 pixels attached to a REOSC 
spectrograph, the size of each pixel being 24 $\mu$m x 24 $\mu$m; one pixel 
corresponds to 0.94$\arcsec$ on the sky. The slit was set in the East-West 
direction and the observations were performed by scanning the slit across 
the objects in the North-South direction in order to get a proper sampling 
of cluster stars. The long slit corresponding to 4.7$\arcmin$ on the sky, 
allowed us to sample regions of the background sky.
We used a grating of 300 grooves mm$^{\rm -1}$, producing an average 
dispersion in the observed region of $\approx$ 140 \AA/mm (3.46 \AA/pixel). 
The spectral coverage was $\approx$ (3600-6800)\,\AA. The slit width was 
4.2$\arcsec$, providing a resolution
(FWHM) of $\approx$ 14\,\AA, as deduced from the comparison lamp lines.

The reduction of the spectra was carried out with the IRAF\footnote{IRAF is 
distributed by the National Optical Astronomy Observatories, which is operated 
by the Association of Universities for Research in Astronomy, Inc., under 
contract with the National Science Foundation} package at the Observatorio 
Astron\'omico de C\'ordoba (Argentina) following the
standard procedures. Summing up, we subtracted the bias and used flat-field 
frames previously combined to correct the frames for high and low spatial 
frequency variations. We also checked the instrumental signature with the 
acquisition of dark frames. Then, we performed the background sky subtraction 
using pixel rows from the same frame, after having cleaned the background sky 
regions from cosmic rays. We controlled that no significant background sky 
residuals were present on the resulting spectra. The cluster spectra, 
which were 
extracted along the slit according to the cluster size and available flux, 
were then wavelength calibrated by fitting observed Cu-Ar-Ne 
comparison lamp spectra with template spectra. The rms errors involved in these
calibrations are on average 0.40 \AA. Finally, extinction correction 
and flux calibrations 
derived from the observed standard stars were applied to the cluster spectra. 
In addition,
cosmic rays on them were eliminated. Table~1 
presents the cluster sample including the designations in different 
catalogues, the equatorial coordinates, the average diameters according to 
\cite{bsdo99} and the averaged signal-to-noise (S/N) ratios of the 
spectra. Typically, three spectra were obtained per cluster.

A colour-colour diagram of the sample is presented in Fig.~1,
where the clusters integrated UBV colours are situated 
among those from a larger sample \citep{bcdsp}
characterizing the whole age range of LMC clusters.

\section{Analysis of the cluster spectra}

Cluster ages were derived by means of two methods: the template 
matching method, in which the observed spectra are compared and matched to 
template spectra with well-determined properties \citep[e.g.][and references 
therein]{pbc02}, and the EW method, in which diagnostic 
diagrams involving the sum of EWs of selected spectral lines were employed 
together with their calibrations with age and metallicity (SP). 
In the first 
method, a high weight is assigned to the overall continuum, 
while in the second method the spectral lines are the relevant observables. 
Both methods rely on the library of star cluster integrated
spectra with well-determined properties, accomplished in various studies 
\citep[e.g.][ and references therein]{ba86,pbc02} and made available through 
the CDS/Vizier catalogue database 
at http://vizier.u-strasbg.fr/cgi-bin/VizieR?-source=III/219 \citep{sab02}.

\subsection{Equivalent width method}

Before measuring EWs in the observed spectra, these  were set to the 
rest-frame according to the Doppler shift of H Balmer lines. 
Next, the spectra 
were normalized at approximately 5870 \AA.  
The flux normalization at this wavelength is meant to
represent the continuum flux around 5870\ \AA, avoiding spectral 
lines eventually
present. In practice, the spectral region around
5870\ \AA\ ($\approx$20\ \AA\ wide) is examined and the normalization 
applied to a nearby
wavelength that is representative of the continuum flux.
 
Spectral fluxes at 3860\,\AA, 4020\,\AA, 4150\,\AA, 4570\,\AA, 4834\,\AA, 
4914\,\AA~ and 6630\,\AA~ were used 
as guidelines to define the continuum according to \cite{ba86}. The 
EWs of H Balmer,  K\,Ca\,II, G band (CH) and Mg\,I (5167 + 5173 + 5184)\,\AA~ 
were measured within the spectral windows defined by \cite{ba86} and using IRAF
task {\it splot}. Boundaries for the K\,Ca\,II, G band (CH), Mg\,I, H$\delta$, 
H$\gamma$ and  H$\beta$ spectral windows are, respectively, (3908-3952) \AA, 
(4284-4318) \AA, (5156-5196) \AA, (4082-4124) \AA, (4318-4364) \AA, and 
(4846-4884) \AA. Such a procedure has been consistently applied, making the 
EWs from integrated spectra safely comparable with those in the well-known 
cluster 
database. Table~2 presents these measurements as well 
as the sum of EWs of 
the three metallic lines ($S_m$) and of the three Balmer lines H$_{\delta}$, 
H$_{\gamma}$ and H$_{\beta}$ ($S_h$). 
Typical errors of $\approx$ 10 \% on individual EW 
measurements were the result of tracing slightly different continua. If
the sums of EWs $S_h$ and $S_m$  are separately used, the EW relative 
errors are 
improved ($\approx7\%$ smaller than the individual EW errors).
$S_m$ and $S_h$ prove to be useful in
the discrimination of old, intermediate-age and young systems 
\citep[][ SP]{r82,dbc99}. 

As a first approach to get cluster ages, Fig.~2 shows the 
cluster sample plotted in the diagnostic 
diagrams defined by SP,  which
are intended to discriminate cluster ages for systems younger than 10\,Gyr,
and metallicities for systems older than 10\,Gyr.
In order to show how sensitive the EW sums are 
on the size of the spatial profile extraction
and on the stochastic effects produced by few bright stars in the integrated
spectrum, we used the clusters {NGC\,1902}
and {SL\,709}. Two spectra were extracted from the spatial profile of 
{NGC\,1902}, one of them sampling a larger cluster extent (resulting a 
smaller $S_h$)
than the other. Besides
the observed integrated spectrum of {SL\,709}, another
spectrum was obtained by subtracting a bright
star superimposed on the cluster bulk profile (resulting a larger
$S_h$). 
Subsequently we only used the {NGC\,1902}
spectrum from its larger extraction, since we
judged that it samples more properly the
cluster population and the  {SL\,709} spectrum with the bright star 
subtracted, which
also should better represent the cluster bulk stellar content.
The linked symbols in the diagrams correspond to two different spectral
extractions from the spatial profiles of  {NGC\,1902} and {SL\,709},
which exemplifies the sensitivity of the integrated spectra on the 
size of the extraction (for {NGC\,1902}) and on the presence of a bright star
superimposed on the cluster bulk profile (for {SL\,709}).
The integrated spectra
of {NGC\,1902} resulted in EW sums which place the cluster
in the same region of the diagrams, whereas for {SL\,709}
a transition occurs between two regions with intermediate ages. 
Indeed, 
{SL\,709} seems to be still further affected by an additional 
superimposed
bright star of a younger age, better representing the cluster bulk
stellar content, as discussed in Sect.~3.2.

The 
clusters were then age-ranked according to the calibrations provided by
SP. Since the EW of each Balmer line is a 
bivalued function of age with a maximum around 300\,Myr, we
used $S_m$ to get a first age estimate using:

\begin{equation}
\log{\rm t(Gyr)}=a_0+a_1.{S_m}+a_2.{S_m}^2,
\end{equation}

\noindent where $a_0=-2.18\pm0.38$, $a_1=0.188\pm0.080$ and $a_2=-0.0030\pm0.0032$.

We then used $S_h$ to get a second age estimate guided by the
previous $S_m$ estimate, since from $S_h$ two solutions are possible:

\begin{equation}
\log{\rm t(Gyr)}=\frac{-b\pm\sqrt{b^2-4.a.(c-S_h)}}{2.a},
\end{equation}

\noindent where $a=-6.35\pm0.18$, $b=-8.56\pm0.35$ and $c=23.32\pm0.20$.

The average of these two estimates is listed in column 5 of Table~3.

\subsection{Template matching method}

The template matching method consists in achieving the best possible match 
between the analyzed cluster spectrum and a template spectrum of known age and 
metallicity. In this process we selected, from the available template spectra,
the ones which minimize the flux residuals, calculated as the normalized 
difference 
(cluster - template)/cluster. Note that differences between cluster and 
template spectra are expected to be found due to variations in the stellar 
composition of the cluster, such as the presence of a relatively bright 
star with particular spectral features or contamination of a field star close
to the direction towards the cluster. 

All 17 clusters in our sample are well represented by blue stellar
populations, according to their spectral properties. For the present sample,
the useful template spectra  are:  Yd (40 Myr),
Ye (45-75 Myr), Yf (100-150 Myr), Yg (200-350 Myr) and Yh (0.5\,Gyr), 
which represent young and intermediate-age populations built from Galactic open
clusters  \citep{pbc02}.

Since the continuum distribution is also affected by reddening, we firstly 
adopted a colour excess $E(B-V)$ for each cluster, taking into account the 
\citet[][ hereafter BH]{bh82} extinction 
maps. Secondly, we corrected the observed spectra accordingly and then we
applied the 
template match method. The results are shown in 
Figs.~6 to \ref{sl709}.

Three clusters ({SL\,237}, {SL\,508} and {SL\,709})
have their spectra affected by a bright star, which was taken into account
according to the following procedure: 
we matched the spectrum of {SL\,237} with a combination of templates
Yd (40\,Myr) and Ye (60\,Myr) and an average spectrum of giant
stars of early M type from \cite{sc92}. The stellar spectrum contributes 
with 35\% of the total light at 5870\,\AA~(see Fig.~\ref{sl237}).
As for {SL\,237}, the spectrum of {SL\,508} shows bands 
redwards of 5000\,\AA, 
characteristic of late spectral types. The same stellar spectrum used
in the spectral matching of {SL\,237} was also used for SL\,508
in combination with template Ye. The star contributes with 15\%
of the integrated light  at 5870\,\AA~(see Fig.~\ref{sl508}). 
The same procedure was applied to {SL\,709}, which presents a 
flat continuum
beyond 5000\,\AA. In the case of {SL\,709}, the spectrum of the 
Carbon star
TT Tau \citep{bsk96}, with  20\% of contribution to the 
total flux at 5870\,\AA, was combined with template Yf (see Fig.~\ref{sl709}).

\section{Adopted ages}

The ages determined by the two methods, together
with estimates from the literature (whenever available), were used to 
get final averaged ages (Table~3).
Their respective errors take into consideration the 
dispersion of the values averaged. The reddening adopted in the template
match method is presented in column 2 of Table~3.
The LMC is optically thin, the average foreground and internal E(B-V) 
colour excesses being 0.06 and 0.06 mag, respectively \citep{dbc01}.
This explains why the derived E(B-V) values turn out to be relatively 
small, being all lower than 0.1 mag.

Twelve of the clusters in our sample are also present in OGLEII sample
\citep{pukaswz99}. \cite{pu00} determined their ages by fitting isochrones 
of Z=0.008 on cluster colour-magnitude diagrams (CMDs) built with OGLEII data 
(see Table~3).
By analyzing  the distribution in the BV photographic CMD
of cluster member stars, \cite{al87} derived age for 
{NGC\,1839}, {NGC\,1870} and {SL\,237} considering 
the position of the 
main sequence turnoff, the position of the brightest blue star and 
the fitting of \cite{mm81} isochrones. Their results are displayed in 
Table~3.
\cite{bmc05} determined the age of {NGC\,1943} by analyzing the 
pulsational period
of its 9 Cepheids and found it consistent with the value obtained from 
isochrone fitting on OGLEII data. The value shown in Table~3 
is an average of their results. 

Within the expected uncertainties, the ages derived in the present work 
agree with those given in the literature.

\section{Age vs. spatial distribution}

\cite{pbg03} carried out observations in the Washington photometric system
for 6 LMC clusters, which increased up to 37 the total sample of young and
intermediate age clusters with
uniform estimates of age and metallicity. These parameters were determined
using CMDs and theoretical isochrones. 
Fig.~3 shows the spatial projected distribution of the clusters in
our sample and the sample of 37 clusters previously studied (upper panel).
Right ascension and declination are relative to the {LMC} center, 
considered
as the position of the cluster {NGC\,1928} ($\alpha$(J2000)=5h20m57s, 
$\delta$(J2000)=$-69^{\circ}28\arcmin41\arcsec$).
The lower panel of this Figure discriminates clusters of different ages
in the whole sample of 53 clusters ({NGC\,1839} is in both samples). 
It can be noticed that the general tendency is for 
the older clusters to lie in the outer disk regions of the galaxy while
the younger ones tend to be located not far from or in the bar. 
This effect is easily observed when the deprojected galactocentric 
distance is plotted against age, as shown in Fig.~4. The
deprojected galactocentric distance was calculated using an inclination
between the outer LMC disk and the plane of the sky of 45$^{\circ}$ and a 
position angle of the line of nodes of 7$^{\circ}$ \citep{lw63}.
This tendency is 
compatible with the findings of \cite{sh02} who derived the {LMC}
star formation history from HST observations of field stars. They present a
detailed comparative analysis of CMDs of the bar
and the disk, concluding that the star formation history is different
in these regions for ages\,$<6$\,Gyr. The star formation rate seems to have
increased around $\approx$2\,Gyr ago for both disk and bar, though it has 
declined more recently
in the disk while remaining roughly constant in the bar
\citep{sh02}. However, cluster formation does not seem to follow star formation
in their detailed histories.

\section{Age vs. integrated colours}

Integrated UBV colours of 624 {LMC} associations and clusters were
obtained by \cite{bcdsp} from photoelectric photometry observations.
We used these data to check the trend of (U-B) and (B-V) with age
according to the present estimates. For clusters with integrated 
photometry, Fig~5 shows
this comparison for the present sample and for the one 
presented in \cite{pbg03}, totalling 53 clusters with uniform
age estimates. The colour gap seen in both (U-B) and (B-V) is a real
feature first identified by \cite{vdb81} and supported by 
larger cluster samples \citep{bcdsp}. The gap is probably
a natural consequence of cluster evolution with increasing 
metallicities towards the present, and epochs of 
reduced cluster formation between $\approx$300\,Myr and $\approx$1\,Gyr 
\citep[e.g. ][]{gcbb95}. At least in the LMC bar, such a period of 
reduced cluster
formation is not observed for the field stars \citep{sh02}.

\section{Conclusions}

We have determined ages of blue LMC star clusters using their integrated
spectra. The results obtained from two independent methods are in 
agreement with the literature estimates for clusters studied in common.

The present sample, added to a set of 
young and intermediate-age clusters with uniformly determined ages, 
was used with the aim of analyzing their spatial distribution. Clusters
closer to the LMC bar tend to be younger than those at the LMC disk.
The colour gap at (U-B)\,$\approx-0.1$ and (B-V)\,$\approx0.5$ noticed 
in studies with larger cluster samples appears to be 
related to the natural cluster evolution, as suggested by the present
analysis. However, the older  groups SWB\,III, SWB\,IVa and SWB\,IVb 
must be observed spectroscopically and in detail by means of CMDs, 
testing also the possibility of decreased cluster formation 
contributing to the gap.

\begin{acknowledgements}
We would like to thank the CASLEO staff 
members and technicians for their kind hospitality and support during the 
observing runs. We are grateful for the use of the CCD and data acquisition 
system at CASLEO, supported under US National Science Foundation (NSF) 
grant AST-90-15827.  We also thank the anonymous referee for helpful
remarks. This work 
was partially supported by the Brazilian institutions CNPq and FAPEMIG, 
and the Argentinian institutions CONICET, Agencia Nacional de 
Promoci\'on Cient\'{\i}fica y Tecnol\'ogica (ANPCyT) and Agencia 
C\'ordoba Ciencia.
\end{acknowledgements}

\clearpage

\begin{figure}
\resizebox{7cm}{!}{\includegraphics{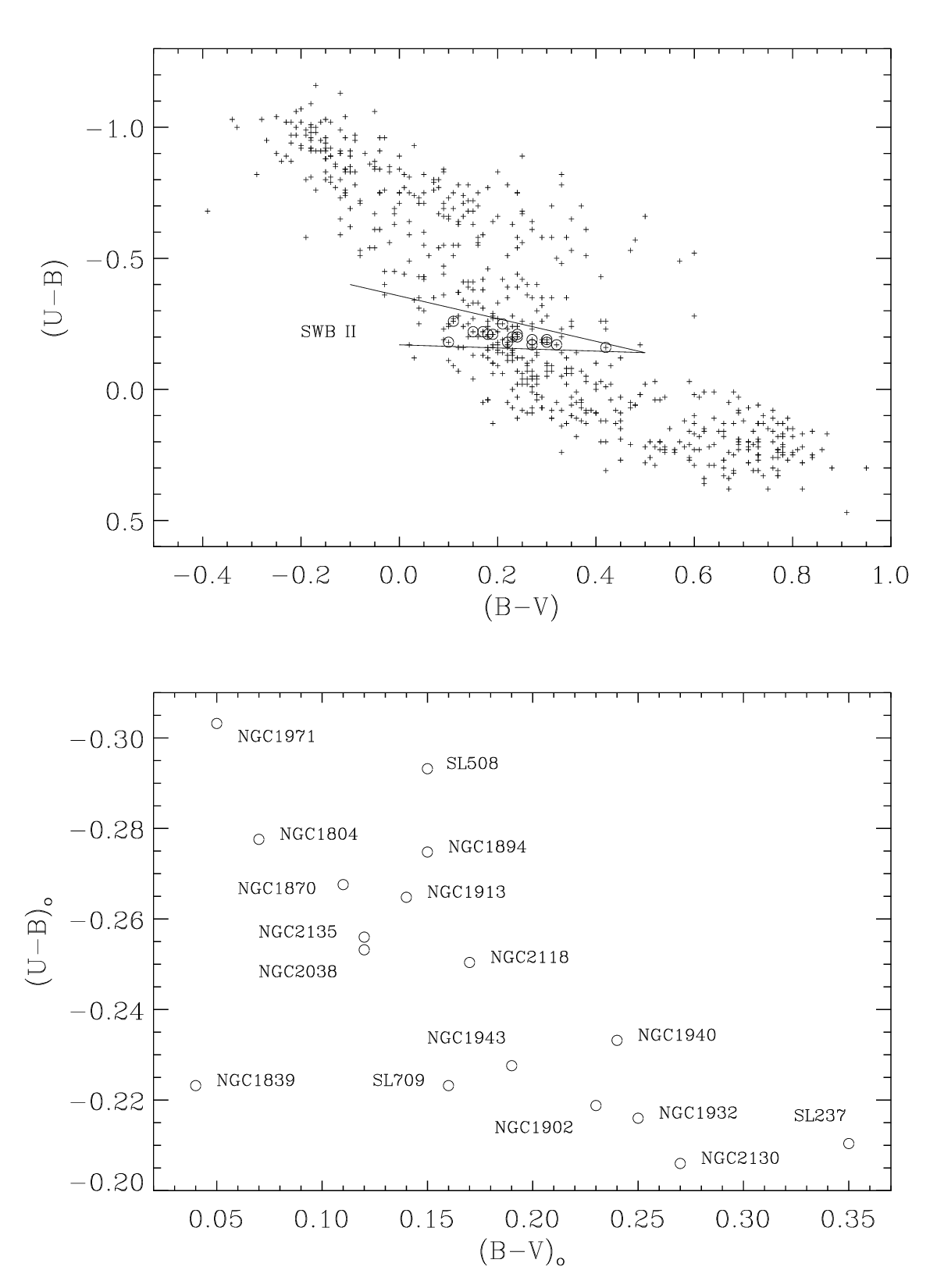}}
\caption{Colour-colour diagrams of the 17 clusters (open circles) and
a larger sample (plus signs) representing the LMC 
cluster system, as observed by means of integrated photometry. 
The lower panel includes only clusters in the present 
sample with colours corrected for reddening from Table~3.}
\end{figure}

\begin{figure}
\resizebox{7cm}{!}{\includegraphics{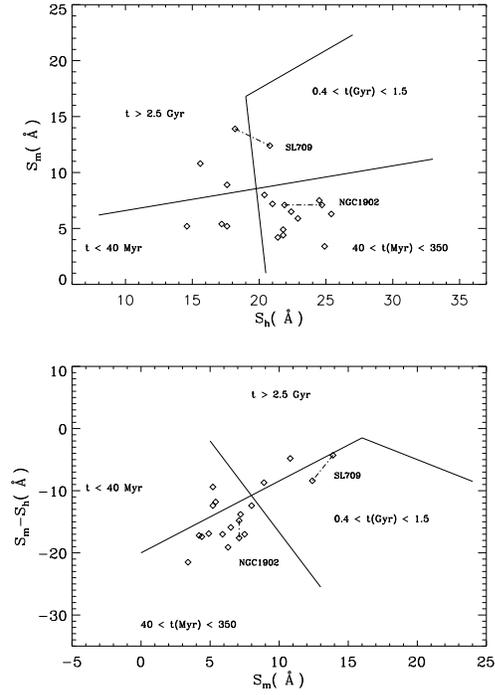}}
\caption{Diagnostic diagrams built from EW sums of H Balmer and metallic
lines for the 17 clusters. Continuous lines divide regions of different
age ranges. The corresponding EW sums for both spectra of {SL\,709}
and {NGC\,1902} are linked.}
\end{figure}

\begin{figure}
\resizebox{7cm}{!}{\includegraphics{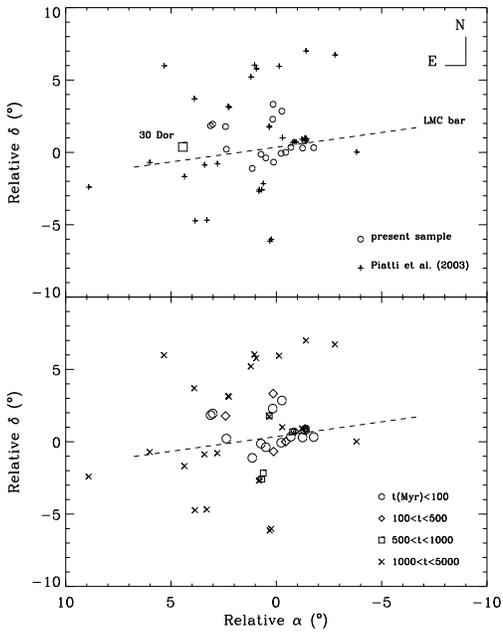}}
\caption{Spatial distribution of {LMC} clusters with ages in 
a homogeneous scale. Upper panel: the present sample and the sample 
analyzed in \cite{pbg03}
are plotted in equatorial coordinates relative to the {LMC} 
center. The positions of the 
bar and {30\,Dor} are indicated. Lower panel: the same data
is plotted with different symbols discriminating different age ranges}
\end{figure}

\begin{figure}
\resizebox{7cm}{!}{\includegraphics{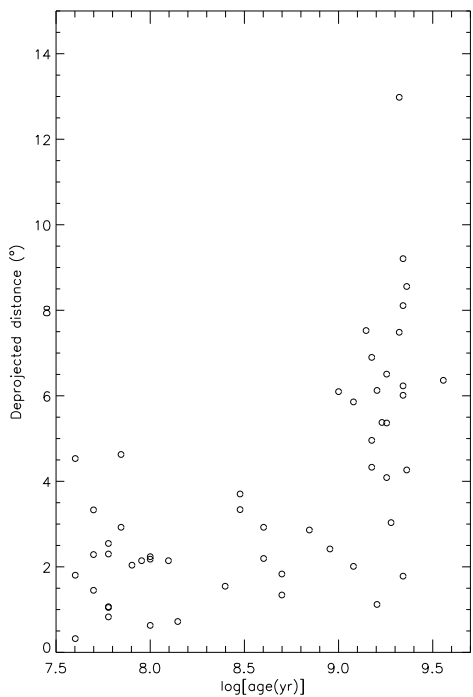}}
\caption{Deprojected distance from the {LMC} center 
(in degrees) as a function of age for the same sample of Fig.~3}
\end{figure}

\begin{figure}
\resizebox{7cm}{!}{\includegraphics{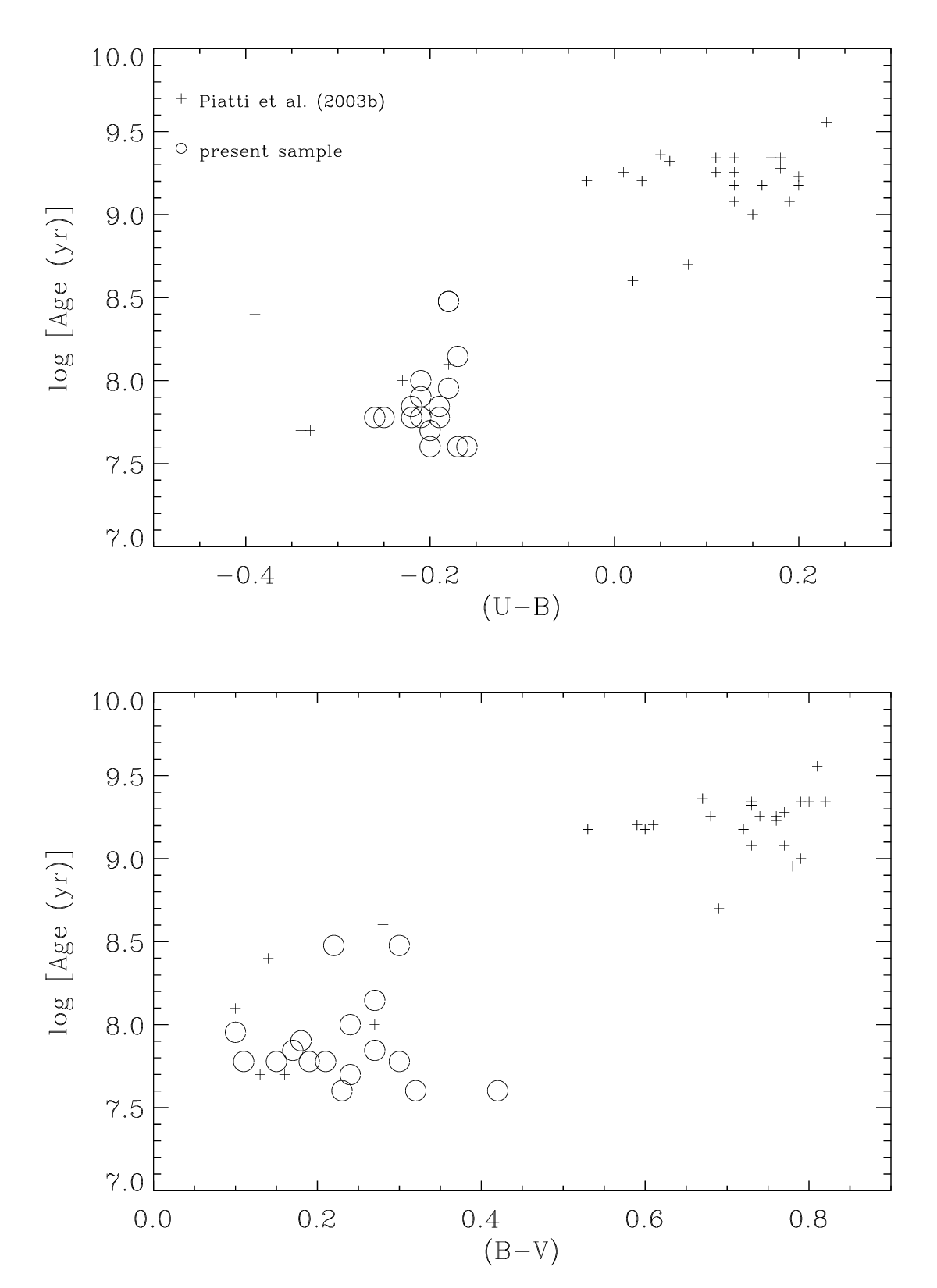}}
\caption{Distribution of cluster age with integrated colours for 
the 17 clusters plus the sample used by Piatti et al. (2003b)}
\end{figure}

\begin{figure}
\resizebox{7cm}{!}{\includegraphics{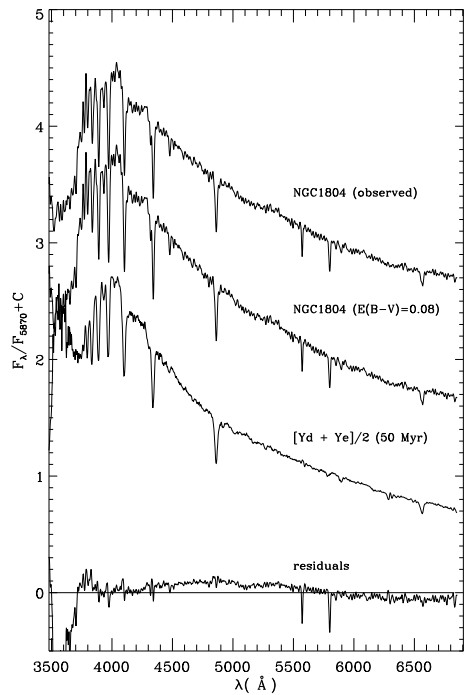}}
\caption{From top to bottom:
the observed integrated spectrum of NGC\,1804, the spectrum corrected
for the adopted reddening $E(B-V)$, the template spectrum which best
matches it, and the flux residuals according to $(F_{cluster}-F_{template})/F_{cluster}$.}
\end{figure}


\begin{figure}
\resizebox{7cm}{!}{\includegraphics{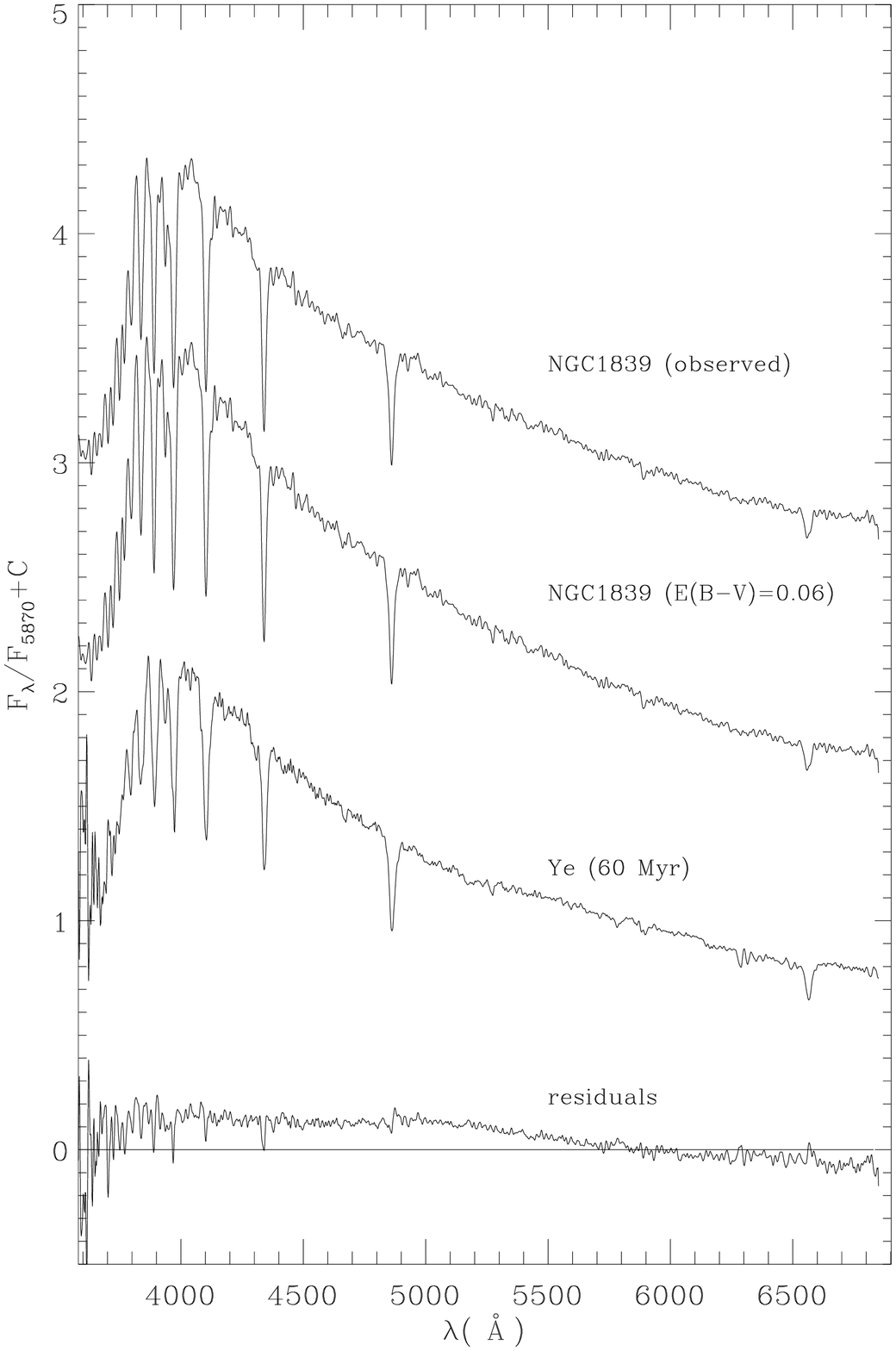}}
\caption{Same as Fig.~6 for NGC\,1839.}
\label{n1839}
\end{figure}

\begin{figure}
  \resizebox{7cm}{!}{\includegraphics{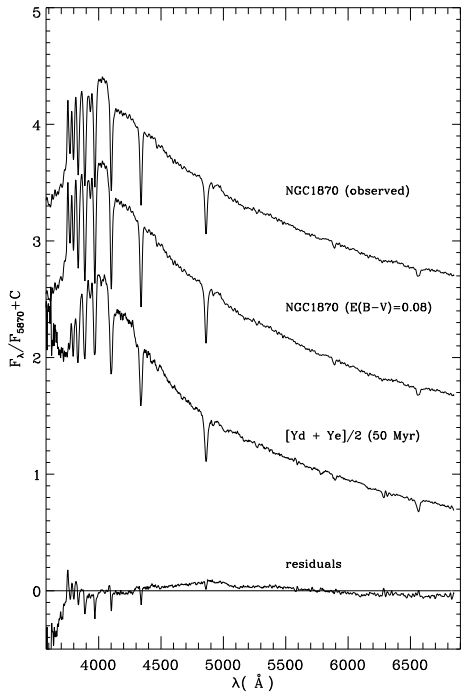}}
\caption{Same as Fig.~6 for NGC\,1870.}
\label{n1870}
\end{figure}

\begin{figure}
\resizebox{7cm}{!}{\includegraphics{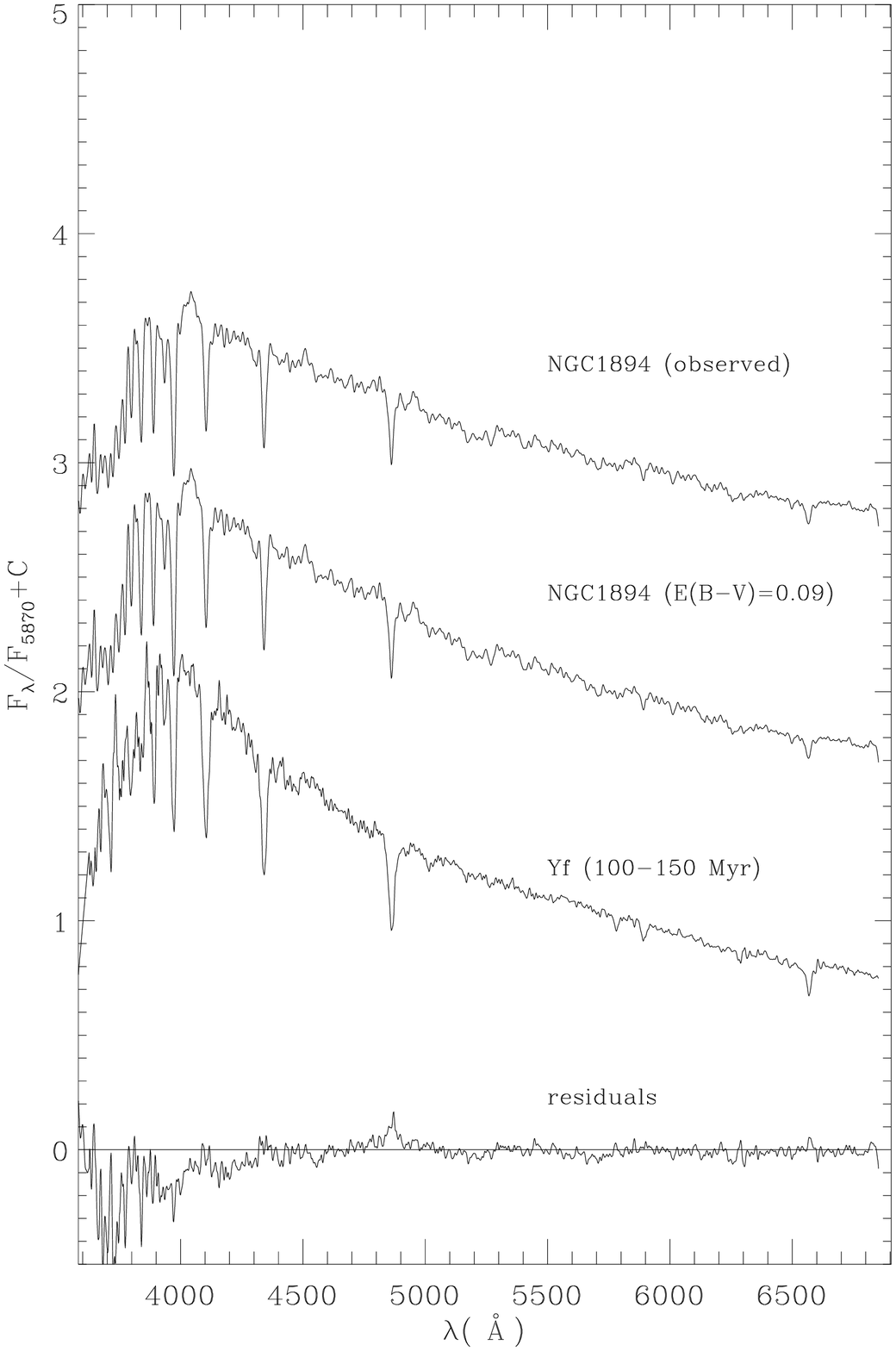}}
\caption{Same as Fig.~6 for NGC\,1894.}
\label{n1894}
\end{figure}

\begin{figure}
\resizebox{7cm}{!}{\includegraphics{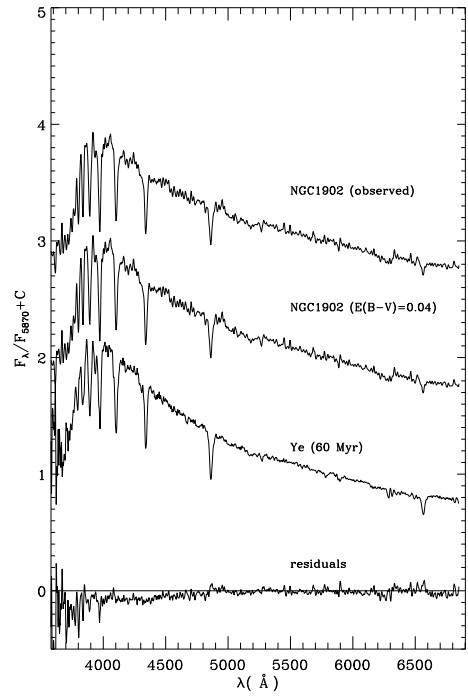}}
\caption{Same as Fig.~6 for NGC\,1902 (larger extraction).}
\label{n1902}
\end{figure}

\begin{figure}
\resizebox{7cm}{!}{\includegraphics{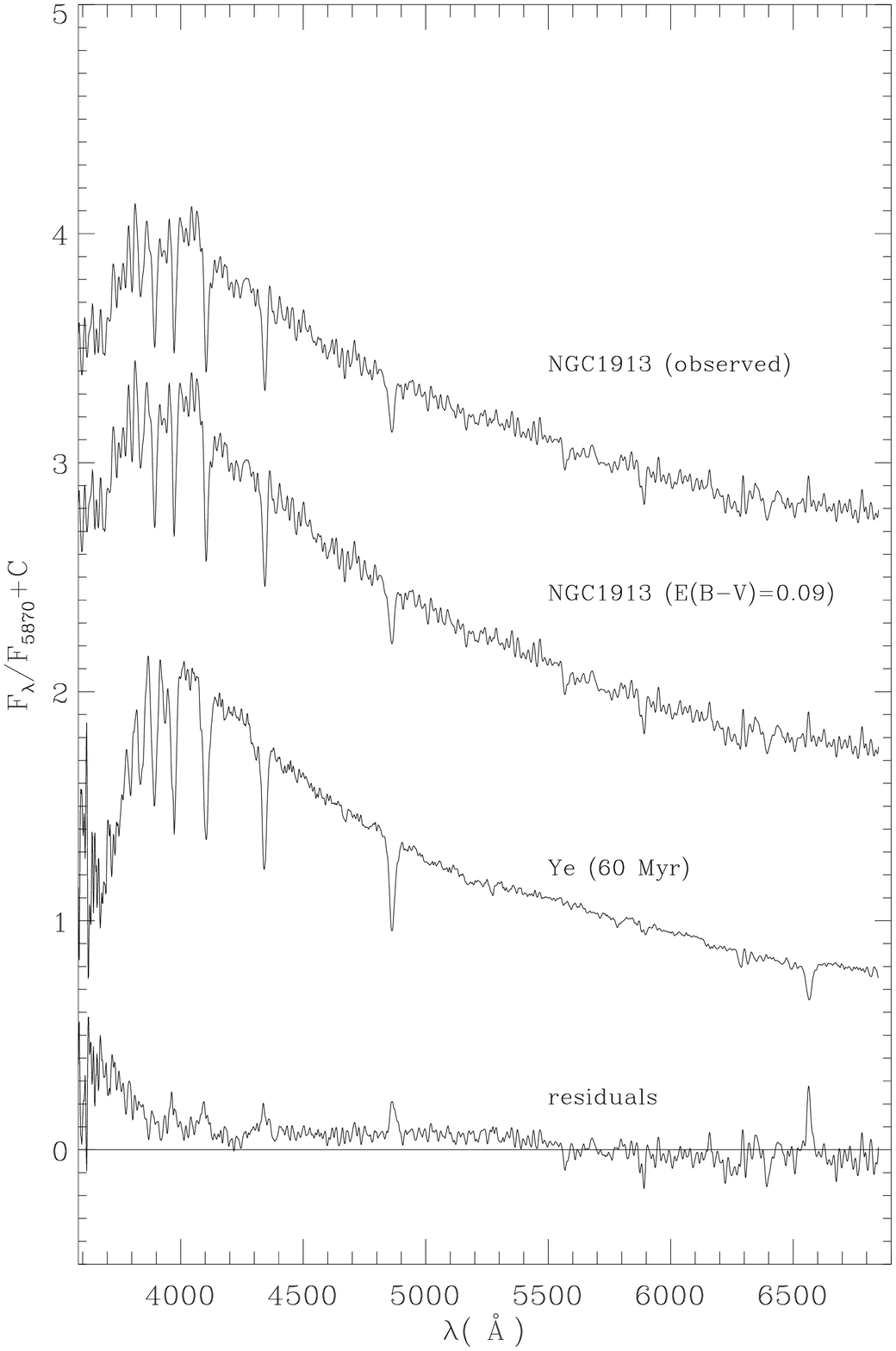}}
\caption{Same as Fig.~6 for NGC\,1913.}
\label{n1913}
\end{figure}

\begin{figure}
\resizebox{7cm}{!}{\includegraphics{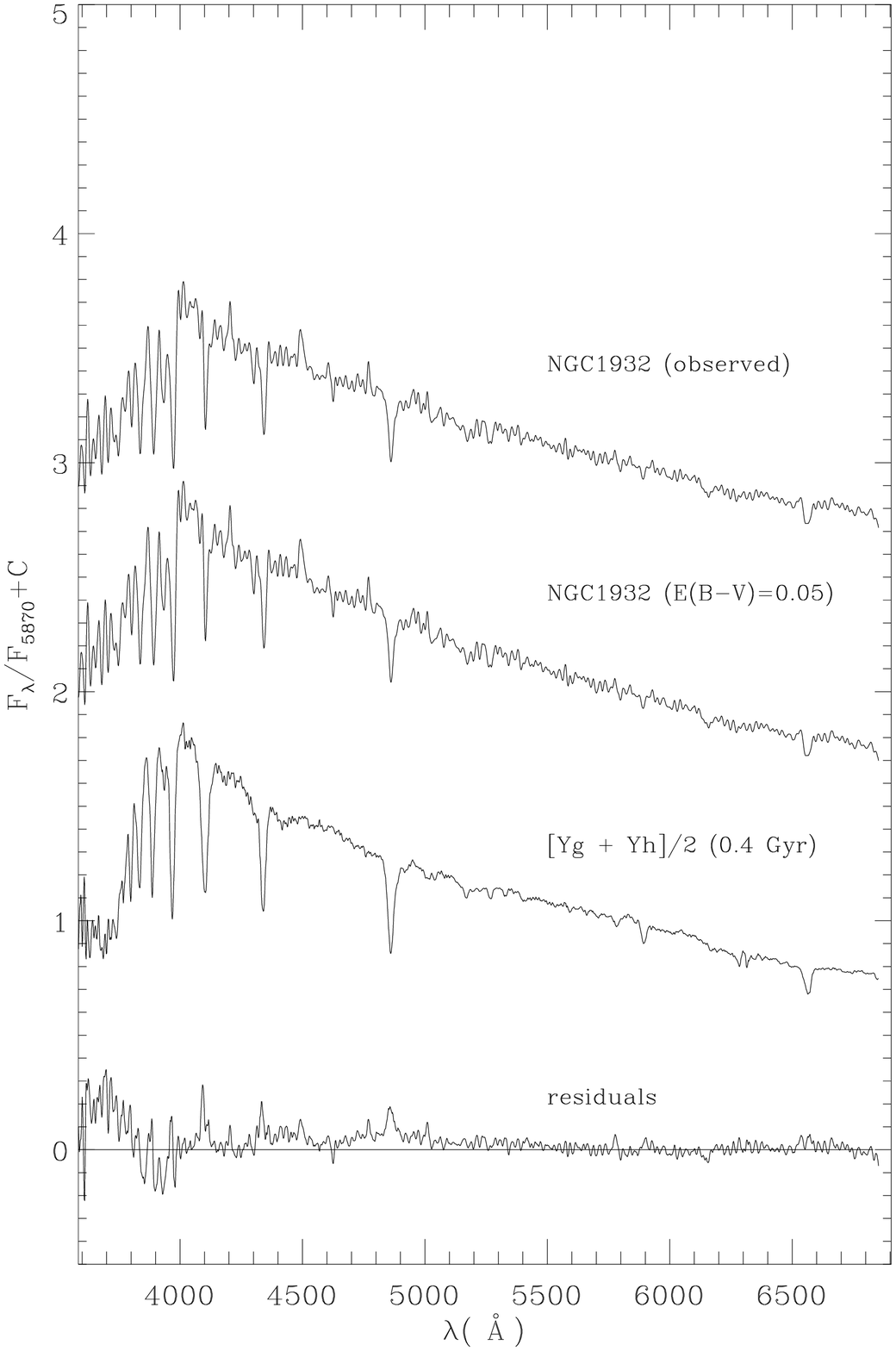}}
\caption{Same as Fig.~6 for NGC\,1932.}
\label{n1932}
\end{figure}

\begin{figure}
\resizebox{7cm}{!}{\includegraphics{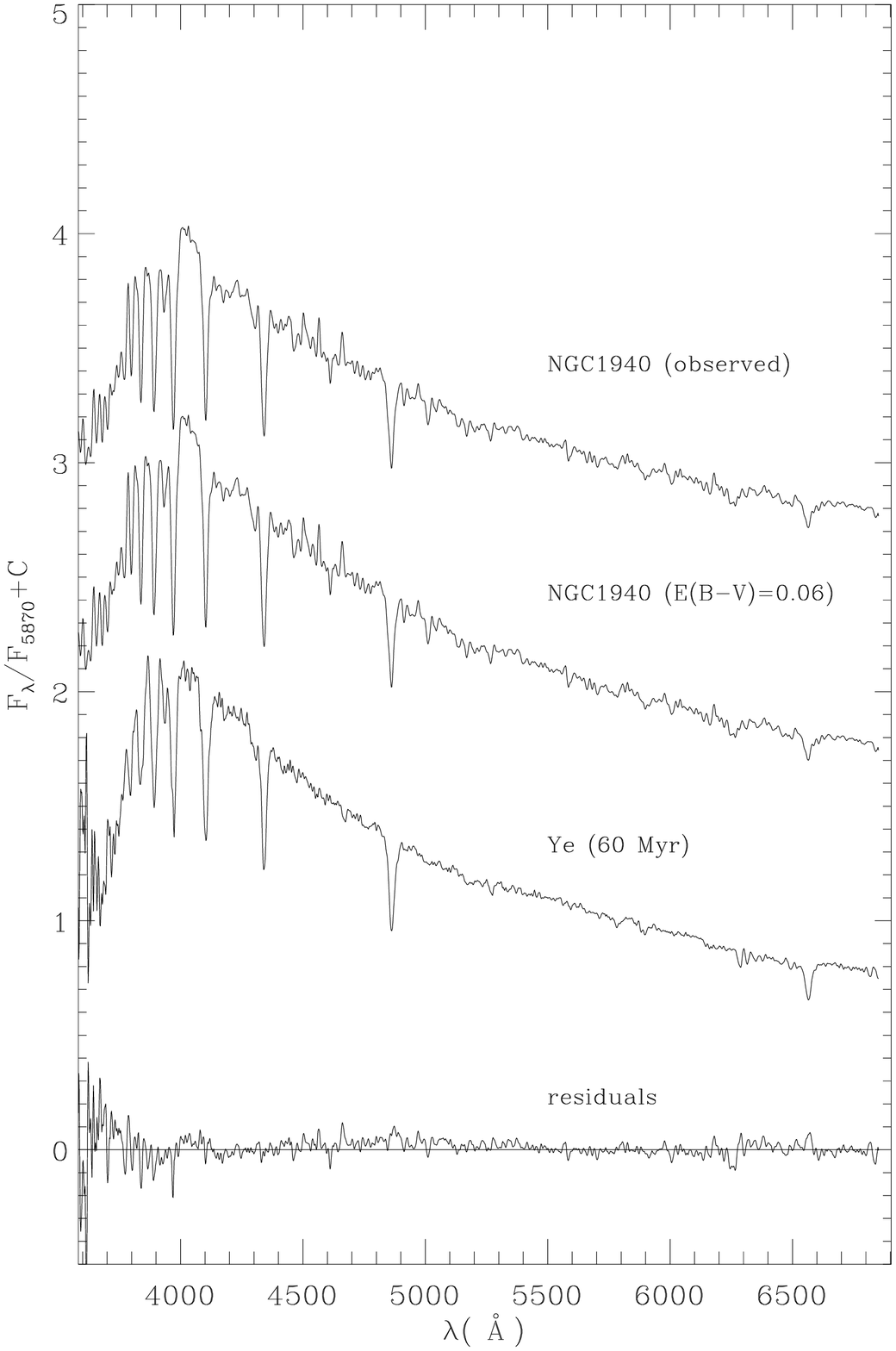}}
\caption{Same as Fig.~6 for NGC\,1940.}
\label{n1940}
\end{figure}

\begin{figure}
\resizebox{7cm}{!}{\includegraphics{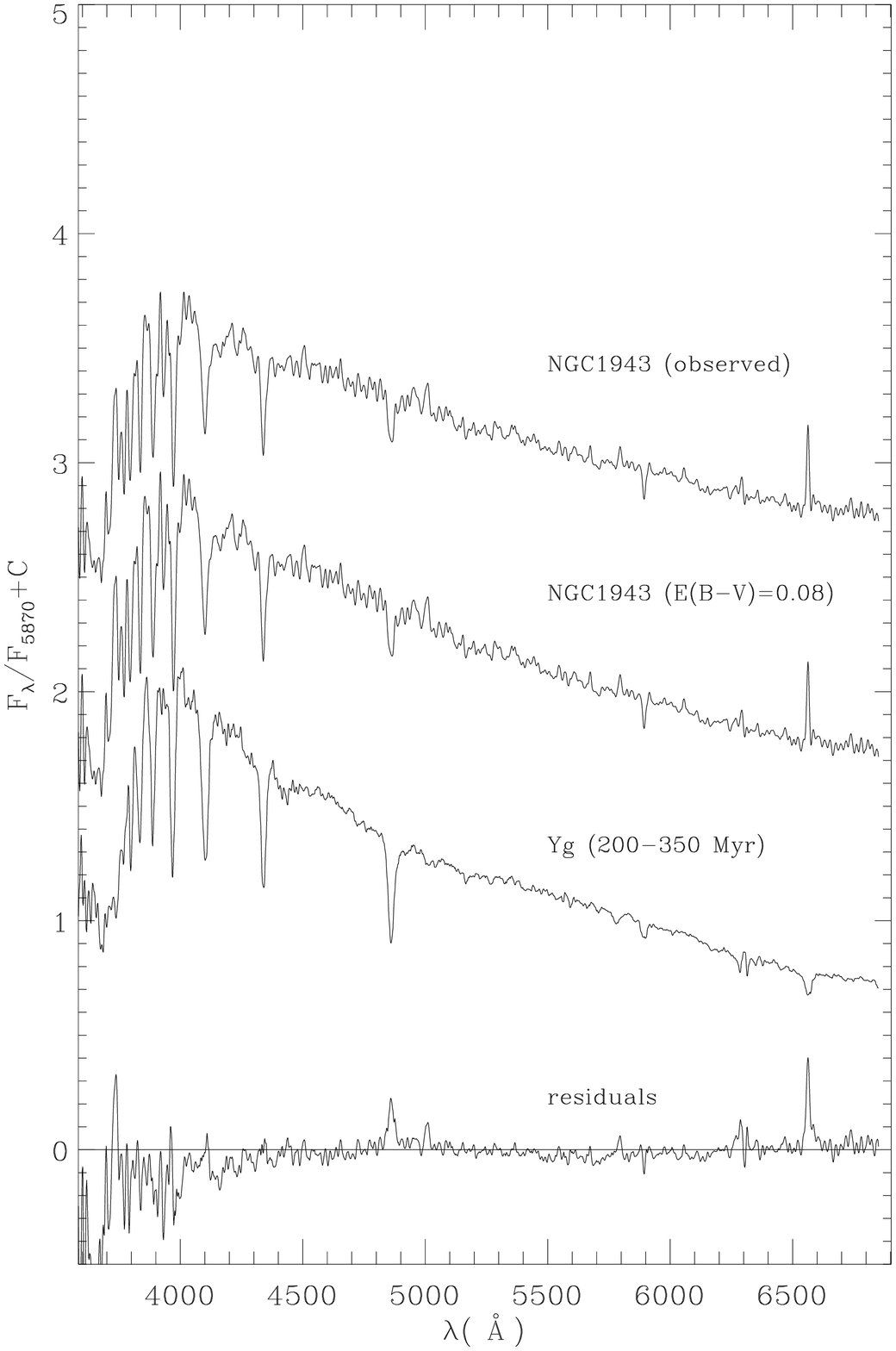}}
\caption{Same as Fig.~6 for NGC\,1943.}
\label{n1943}
\end{figure}

\begin{figure}
\resizebox{7cm}{!}{\includegraphics{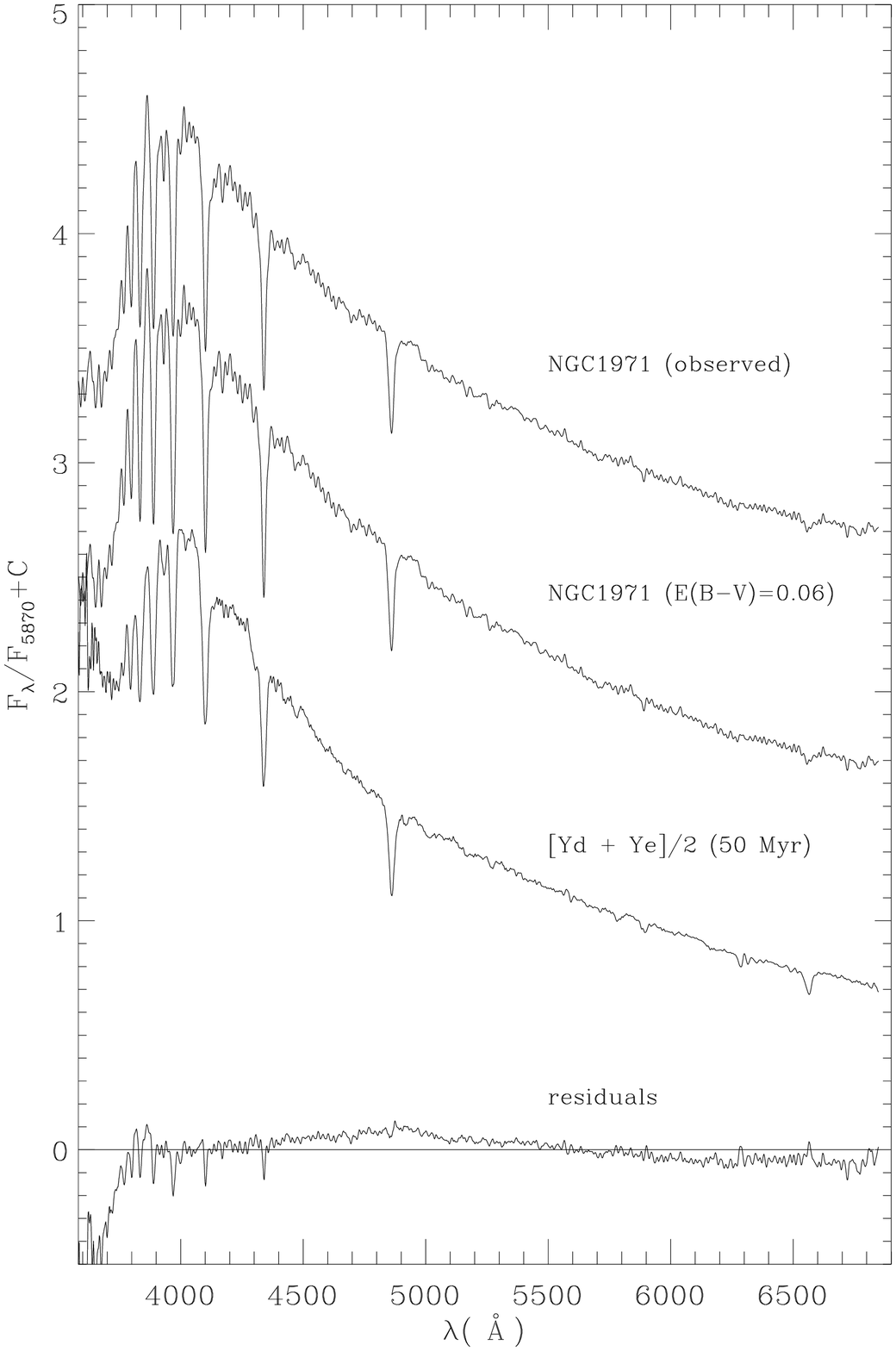}}
\caption{Same as Fig.~6 for NGC\,1971.}
\label{n1971}
\end{figure}

\begin{figure}
\resizebox{7cm}{!}{\includegraphics{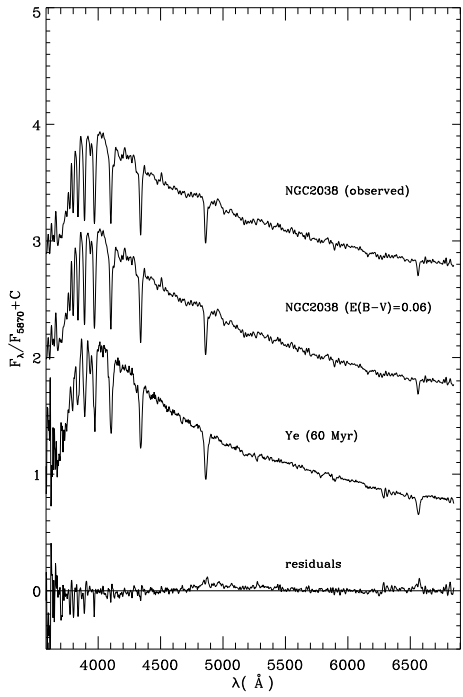}}
\caption{Same as Fig.~6 for NGC\,2038.}
\label{n2038}
\end{figure}

\clearpage

\begin{figure}
\resizebox{7cm}{!}{\includegraphics{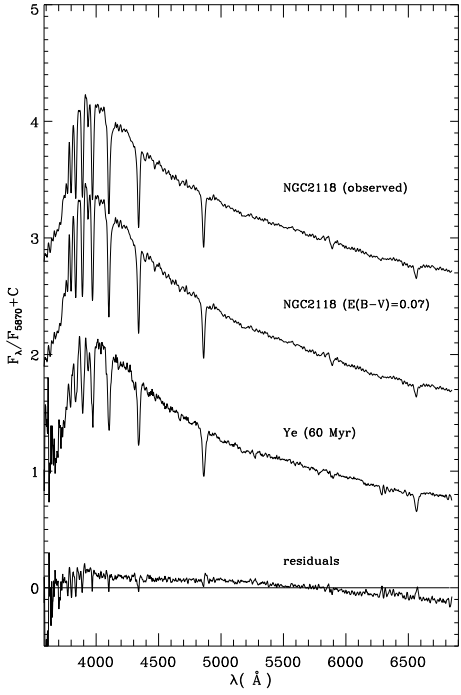}}
\caption{Same as Fig.~6 for NGC\,2118.}
\label{n2118}
\end{figure}

\begin{figure}
\resizebox{7cm}{!}{\includegraphics{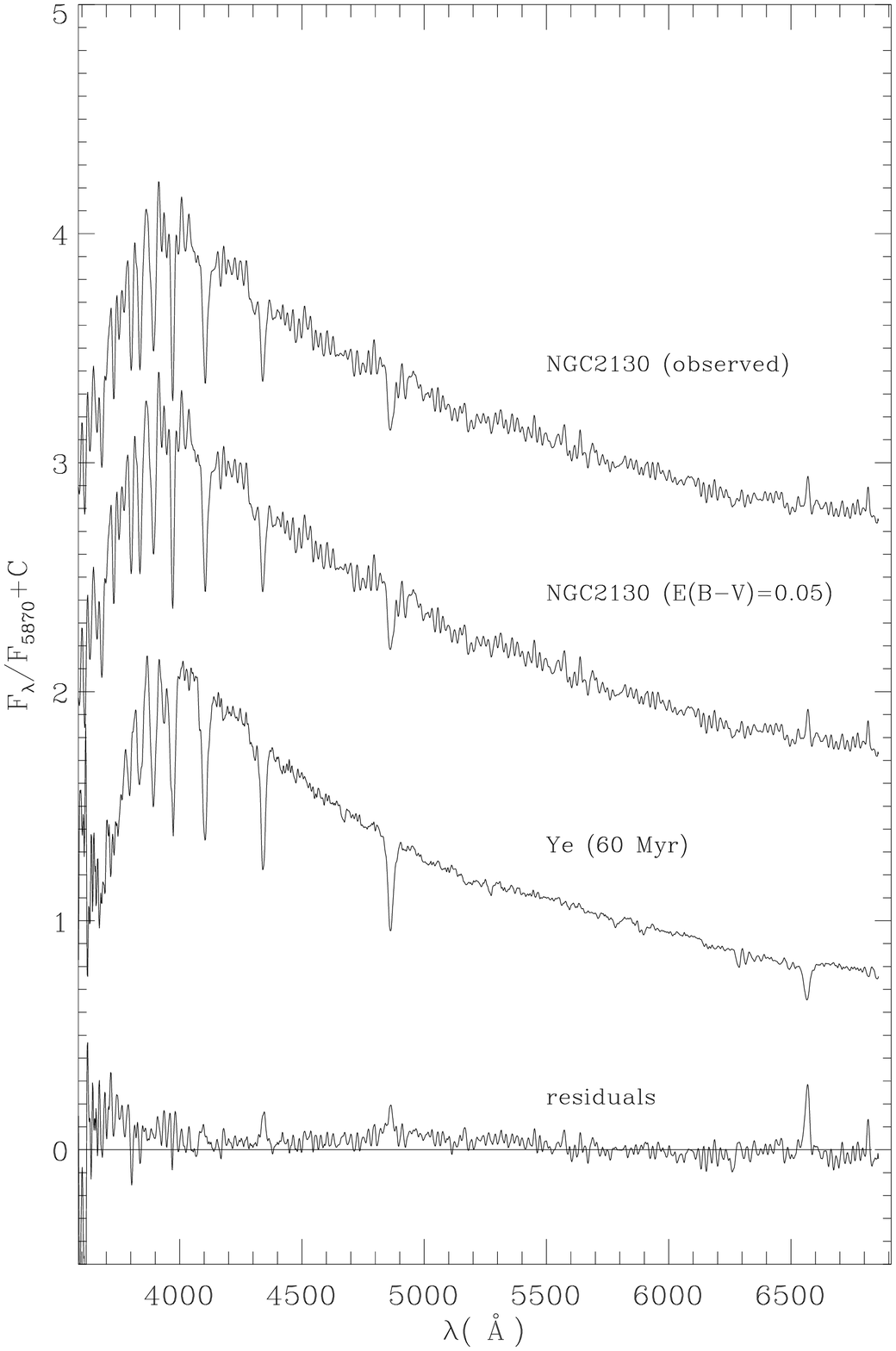}}
\caption{Same as Fig.~6 for NGC\,2130.}
\label{n2130}
\end{figure}

\begin{figure}
\resizebox{7cm}{!}{\includegraphics{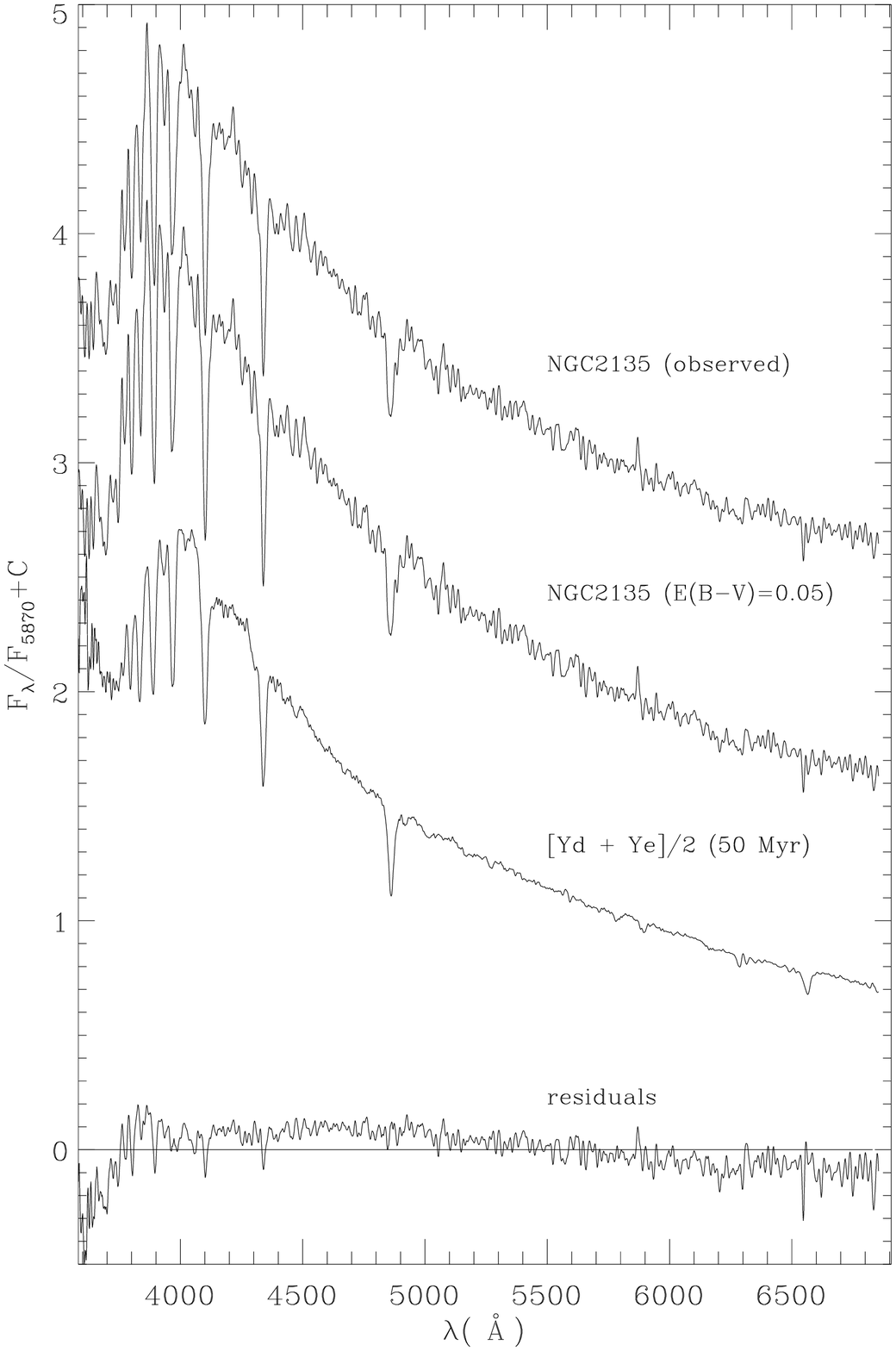}}
\caption{Same as Fig.~6 for NGC\,2135.}
\label{n2135}
\end{figure}

\begin{figure}
\resizebox{7cm}{!}{\includegraphics{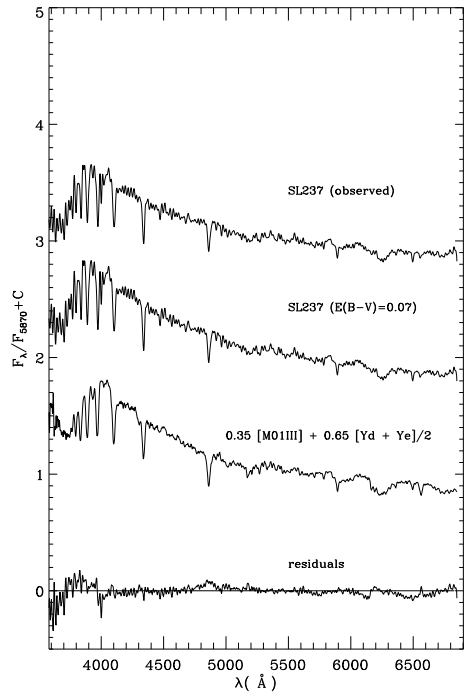}}
\caption{Same as Fig.~6 for SL\,237. The matched spectrum is
a combination of templates Ye (40\,Myr) and Yd (60\,Myr) and an
average spectrum of early M giants \citep{sc92}. The flux fraction 
at 5870\AA\ is indicated.}
\label{sl237}
\end{figure}

\begin{figure}
\resizebox{7cm}{!}{\includegraphics{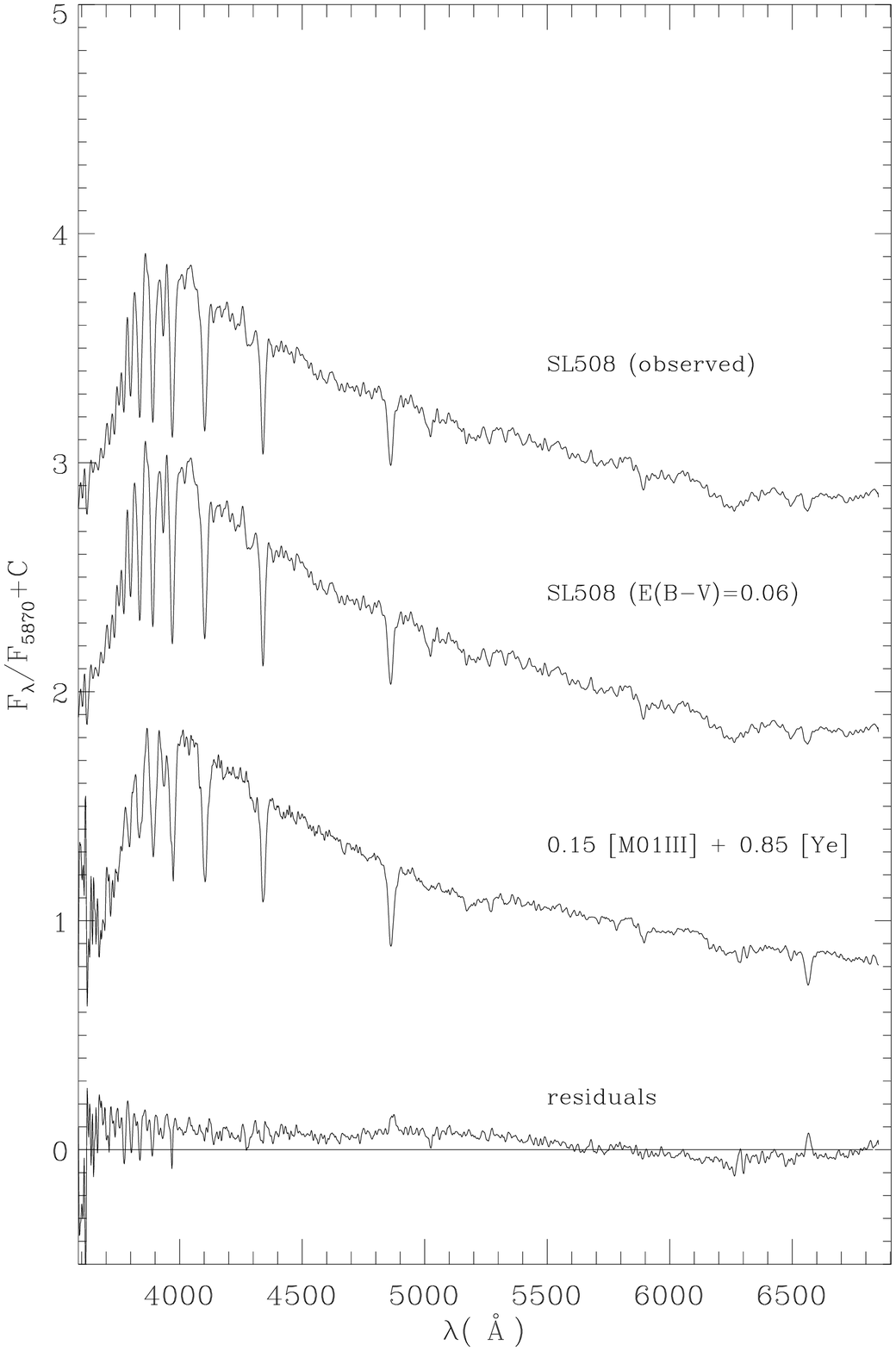}}
\caption{Same as Fig.~6 for SL\,508. The matched spectrum is
a combination of template Ye (40\,Myr) and an
average spectrum of early M giants \citep{sc92}. The flux fraction 
at 5870\AA\ is indicated.}
\label{sl508}
\end{figure}

\begin{figure}
\resizebox{7cm}{!}{\includegraphics{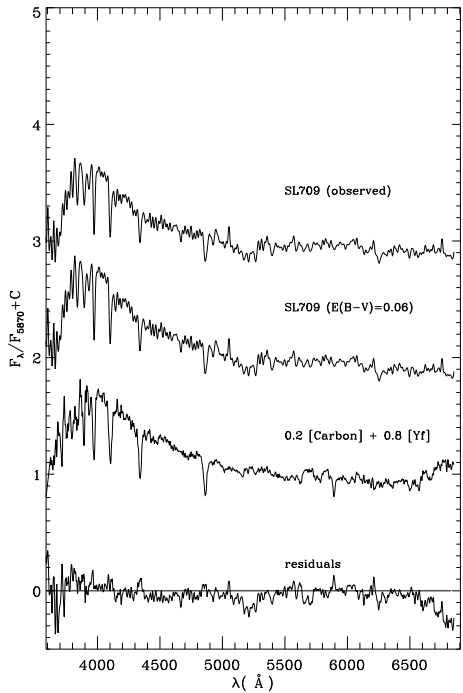}}
\caption{Same as Fig.~6 for SL\,709. The matched spectrum is
a combination of template Yf (100-150\,Myr) and the spectrum of a Carbon
star (TT Tau from \cite{bsk96}). The flux fraction 
at 5870\AA\ is indicated.}
\label{sl709}
\end{figure}

\clearpage
{\onecolumn
\begin{table}
\label{obs}
\caption{The sample clusters}
\begin{tabular}{lcccc}
\hline
Name$^a$  &  $\alpha_{\rm 2000}$ & $\delta_{\rm 2000}$ & D$^b$ & S/N\\
      &  (h:m:s)  & ($\degr$:$\arcmin$:$\arcsec$) & ($\arcmin$) &  \\
\hline
{NGC\,1804, SL\,172, ESO\,56-SC\,46, LMC\_OGLE\,8}              & 5:01:04 &-69:04:57 & 0.90  & 35 \\
{NGC\,1839, SL\,226, ESO\,56-SC\,63, LMC\_OGLE\,93} & 5:06:02 &-68:37:36 & 1.60  & 50 \\
{SL\,237, LMC\_OGLE\,116} & 5:06:57 &-69:08:51 & 1.00  & 30 \\
{NGC\,1870, SL\,317, ESO\,56-SC\,81, LMC\_OGLE\,235} & 5:13:10 &-69:07:01 & 1.05  & 95 \\
{NGC\,1894, SL\,344, ESO\,56-SC\,89, BRHT\,8a, LMC\_OGLE\,278} & 5:15:51 &-69:28:09 & 1.30  & 45 \\
{NGC\,1902, SL\,367, ESO\,85-SC\,66, KMHK\,734} & 5:18:18 &-66:37:38 & 1.70  & 35 \\
{NGC\,1913, SL\,373, ESO\,56-SC\,97, BRHT\,10a, LMC\_OGLE\,330} & 5:18:21 &-69:32:31 & 1.20  & 25 \\
{NGC\,1932, SL\,420, ESO\,85-SC\,77, KMHK\,825} & 5:22:27 &-66:09:09 & 1.30  & 30 \\
{NGC\,1943, SL\,430, ESO\,56-SC\,114, LMC\_OGLE\,411} & 5:22:29 &-70:09:17 & 1.05  & 30 \\
{NGC\,1940, SL\,427, ESO\,85-SC\,78, KMHK\,834} & 5:22:43 &-67:11:10 & 1.20  & 35 \\
{NGC\,1971, SL\,481, ESO\,56-SC\,128, BRHT\,12a, LMC\_OGLE\,480} & 5:26:45 &-69:51:07 & 1.03 & 50 \\
{SL\,508, LMC\_OGLE\,518} & 5:29:20 &-69:35:56 & 1.00  & 45 \\
{NGC\,2038, SL\,590, ESO\,56-SC\,158, KMHK\,1110, LMC\_OGLE\,607} & 5:34:41&-70:33:39 & 1.55 & 45 \\
{SL\,709, BM\,63, KMHK\,1350} & 5:46:15 &-67:34:06 & 1.03  & 25 \\
{NGC\,2118, SL\,717, ESO\,57-SC\,39, KMHK\,1380} & 5:47:39 &-69:07:54 & 1.60  & 60 \\
{NGC\,2130, SL\,758, ESO\,86-SC\,37, KMHK\,1476} & 5:52:23 &-67:20:02 & 1.35  & 25 \\
{NGC\,2135, SL\,765, ESO\,86-SC\,39, BM\,151, KMHK\,1496} & 5:53:35 &-67:25:40 & 1.50  & 20 \\
\hline
\end{tabular}

$^a$ Cluster identifications are from 
Lauberts (1982, ESO), Shapley \& Lindsay (1963, SL), 
Bhatia et al. (1991, BRHT), Kontizas et al. (1990, KMHK),
Bhatia \& MacGillivray (1989, BM), Pietrzy\'nski et al. (1999, LMC\_OGLE).

$^b$ Average diameter according to Bica et al. (1999).
\end{table}

\begin{table}
\label{ews}
\caption{Equivalent widths (\AA)}
\begin{tabular}{lcccccccc}
\hline
Feature &   K\,Ca\,II  & H$_\delta$& G\,band (CH)& H$_\gamma$&H$_\beta$&Mg\,I & $S_h$ & $S_m$\\
Windows (\AA) &3908 &4082&4284       &4318  &4846&5156&&\\
              &-3952&-4124 &-4318    &-4364  &-4884&-5196&&\\
\hline
Cluster & &&&&&&   & \\
{NGC\,1804}    &   1.9  &    7.7  &   1.1   &     8.1   &   6.0  &  1.4& 21.8 & 4.4 \\  
{NGC\,1839}    &   4.7  &    8.6  &   1.7   &     8.7   &   7.2  &  1.1& 24.5 & 7.5 \\  
{SL\,237}    &   0.7  &    6.5  &   1.3   &     6.4   &   4.3  &  3.4& 17.2 & 5.4 \\  
{NGC\,1870}    &   2.5  &    7.8  &   0.2   &     7.0   &   6.6  &  1.5& 21.4 & 4.2 \\  
{NGC\,1894}    &   4.5  &    6.8  &   1.9   &     6.2   &   4.6  &  2.5& 17.6 & 8.9 \\  
{NGC\,1902}$^a$    &   2.9  &    7.6  &   2.0   &     8.4   &   5.9  &  2.2& 21.9 & 7.1 \\  
{NGC\,1902}    &   1.9  &    9.0  &   2.5   &     9.6   &   6.1  &  2.7& 24.7 & 7.1 \\  
{NGC\,1913}    &   2.4  &    5.7  &   0.8   &     5.0   &   3.9  &  2.0& 14.6 & 5.2 \\  
{NGC\,1932}    &   6.4  &    5.2  &   2.2   &     5.5   &   4.9  &  2.2& 15.6 & 10.8 \\  
{NGC\,1943}    &   4.7  &    8.9  &   1.3   &     7.5   &   4.0  &  2.0& 20.4 & 8.0 \\  
{NGC\,1940}    &   3.6  &    8.2  &   0.7   &     7.9   &   6.3  &  2.2& 22.4 & 6.5 \\  
{NGC\,1971}    &   3.3  &    8.3  &   1.1   &     8.3   &   6.3  &  1.5& 22.9 & 5.9 \\  
{SL\,508 }    &   3.4  &    8.5  &   1.1   &     7.0   &   5.5  &  2.7& 21.0 & 7.2 \\  
{NGC\,2038}    &   1.9  &    8.3  &   0.9   &     7.8   &   5.7  &  2.1& 21.8 & 4.9 \\  
{SL\,709 }    &   3.2  &    3.9  &   3.7   &     8.8   &   5.5  &  7.0& 18.2 & 13.9 \\  
{SL\,709}$^b$ &   3.2  &    7.1  &   3.0   &     6.9   &   6.8  &  6.2& 20.8 & 12.4 \\  
{NGC\,2118}    &   1.3  &    8.9  &   0.8   &     8.8   &   7.2  &  1.3& 24.9 & 3.4 \\  
{NGC\,2130}    &   1.3  &    7.4  &   2.2   &     5.1   &   5.1  &  1.7& 17.6 & 5.2 \\  
{NGC\,2135}    &   2.8  &    9.7  &   1.9   &     9.1   &   6.6  &  1.6& 25.4 & 6.3 \\  
\hline
\end{tabular}

$^a$ Larger areal extraction than that for {NGC\,1902}.

$^b$ Bright star subtracted from {SL\,709} integrated spectrum.
\end{table}
}

\clearpage
\begin{table}
\label{param}
\caption{Cluster parameters}
\begin{tabular}{lccrccc}
\hline
Cluster &$E(B-V)$ & $t_{\rm literature}$  &  Ref. &   $t_{\rm Sh,Sm}$& $t_{\rm template}$ &  $t_{\rm adopted}$ \\
        &       &(Gyr)   &       & (Gyr)&  (Gyr)     &    (Gyr)  \\
\hline
{NGC\,1804}    &0.08& $0.08\pm0.01$& 1 & $0.035\pm0.004$ &$0.05\pm0.01$ & $0.06\pm0.02$\\
{NGC\,1839}    &0.06 & $0.10\pm0.01$ & 1 & $0.09\pm0.02$& $0.06$&$0.09\pm0.03$\\ 
             &          & $0.125\pm0.025$& 4 &              &             &             \\
{SL\,237}     &0.07 & $0.038\pm0.004$& 1 & $0.04\pm0.02$&$0.05\pm0.01$&$0.04\pm0.02$\\ 
             &          & $0.027\pm0.009$& 2 &              &             &             \\
{NGC\,1870}    &0.08 & $0.09\pm0.01$& 1 & $0.033\pm0.004$&$0.05\pm0.01$&$0.06\pm0.03$\\ 
             &          & $0.07\pm0.03$& 2 &              &             &             \\
{NGC\,1894}    &0.09 & $0.071\pm0.008$& 1 & $0.10\pm0.08$&$0.13\pm0.03$&$0.10\pm0.03$\\ 
{NGC\,1902}    &0.04 & $-          $& & $0.07\pm0.03$&$0.06$&$0.07\pm0.03$\\
{NGC\,1913}    &0.09 & $0.024\pm0.002$& 1 & $0.03\pm0.02$&$0.06$&$0.04\pm0.02$\\
{NGC\,1932}    &0.05 & $-          $& & $0.2\pm0.2$ &$0.4\pm0.2$&$0.3\pm0.2$\\
{NGC\,1943}   &0.08 & $0.14\pm0.02$& 1 &$0.08\pm0.06$&$0.28\pm0.08$&$0.14\pm0.06$\\
             &          & $0.10\pm0.01$& 3 &              &        &      \\
{NGC\,1940}    &0.06 & $-          $& & $0.06\pm0.02$&$0.06$&$0.06\pm0.02$\\
{NGC\,1971}    &0.06 & $0.10\pm0.01$& 1 & $0.05\pm0.01$&$0.05\pm0.01$&$0.06\pm0.02$\\
{SL\,508}     &0.06 & $0.10\pm0.01$& 1 & $0.07\pm0.04$&$0.06$&$0.07\pm0.04$\\
{NGC\,2038}    &0.06 & $0.13\pm0.02$& 1 & $0.039\pm0.008$&$0.06$&$0.08\pm0.05$\\ 
{SL\,709} &0.06 & $-          $& & $0.3\pm0.2$&$0.13\pm0.03$&$0.3\pm0.2$\\
{NGC\,2118}    &0.07 & $-          $& & $0.05\pm0.02$&$0.06$&$0.05\pm0.02$\\
{NGC\,2130}    &0.05 & $-          $& & $0.03\pm0.02$&$0.06$&$0.04\pm0.02$\\ 
{NGC\,2135}    &0.05 & $-          $& & $0.085\pm0.008$&$0.05\pm0.01$&$0.07\pm0.02$\\ 
\hline
\end{tabular}

References: (1) Pietrzy\'nski \& Udalski (2000); (2) Alcaino \& Liller (1987); 
(3) Bono et al. (2005); (4) Piatti et al. (2003a).
\end{table}

\clearpage
\end{document}